\documentclass[12pt,final]{iopart}
\usepackage{amssymb}
\usepackage{graphicx}
\usepackage[breaklinks=true,colorlinks=true,linkcolor=blue,urlcolor=blue,citecolor=blue]{hyperref}
\usepackage{graphicx}% Include figure files
\usepackage{dcolumn}% Align table columns on decimal point

\begin{document}

\title{Electronic Structure and Transport in Graphene/Haeckelite
       Hybrids: An {\em Ab Initio} Study}

% \author{Zhen Zhu}
% \address{Physics and Astronomy Department,
%              Michigan State University,
%              East Lansing, Michigan 48824, USA}
%
% \author{Zacharias G. Fthenakis}
% \address{Physics and Astronomy Department,
%              Michigan State University,
%              East Lansing, Michigan 48824, USA}
%
% \author{David Tom\'{a}nek}
% \address{Physics and Astronomy Department,
%              Michigan State University,
%              East Lansing, Michigan 48824, USA}
% [E-mail: ]
% \ead{tomanek@pa.msu.edu}

\author{Zhen Zhu, Zacharias G. Fthenakis and David Tom\'{a}nek}
\address{Physics and Astronomy Department,
             Michigan State University,
             East Lansing, Michigan 48824, USA}
\ead{tomanek@pa.msu.edu}%

%\date{ Received }
\date{\today}


\begin{abstract}
We combine {\em ab initio} density functional theory (DFT)
structural studies with DFT-based nonequilibrium Green function
calculations to investigate how the presence of non-hexagonal
rings affects electronic transport in graphitic structures. We
find that infinite monolayers, finite-width nanoribbons and
nanotubes formed of 5-8 haeckelite with only 5- and 8-membered
rings are generally more conductive than their graphene-based
counterparts. Presence of haeckelite defect lines in the perfect
graphitic structure, a model of grain boundaries in CVD-grown
graphene, increases the electronic conductivity and renders it
highly anisotropic.
%In the present work we study the structural, electronic and
%transport properties of haeckelite with 5- and 8-membered rings
%(5-8 haeckelite) and its derivatives, including 1D haeckelite
%nanoribbons, nanotubes and 2D haeckelite and graphene hybrid
%structures. We find the introduction of 5- and 8-membered rings
%could turn semi-metallic graphene to metallic 5-8 haeckelite, in
%which the non-zero electronic density of states at Fermi energy
%could greatly improve the electrical transport properties. Quantum
%confinement in 1D systems could induce a small band gap in
%selected nanoribbons and nanotubes. In graphene/haeckelite hybrid
%structures, the presence of 5-8 defect lines makes the system
%highly anisotropic.
\end{abstract}

%---------------------------------------------------------------------
\maketitle
%---------------------------------------------------------------------

% If in two-column mode, this environment will change to single-column
% format so that long equations can be displayed. Use sparingly.
%\begin{widetext}
% put long equation here
%\end{widetext}

%\section{Outline}\label{outline}

%==================================================================
\section{Introduction}
%==================================================================

Graphene is a unique 2D material that combines extraordinarily
high electrical and thermal
conductivity~\cite{{Novoselov04},{graphene-el-mobility08}} with
mechanical strength, flexibility, thermal and chemical stability.
Interest in this system increased significantly after a successful
mechanical exfoliation using a Scotch tape has been
reported~\cite{Novoselov2005} that yielded large, defect-free
samples. As a scalable alternative to the ``Scotch tape''
exfoliation technique, chemical vapor deposition (CVD) is commonly
being used now to form graphene monolayers on metal substrates
including
Cu~\cite{{KimNat2009},{ReinaNL2009},{KSKimgraphene09},%
{Ruoff-graphene-on-copper09}}. The quality of CVD-grown films is
inferior to those produced by exfoliation, since gas phase
deposition leads to simultaneous growth of graphene flakes that
eventually interconnect, forming grain boundaries with a
defective, haeckelite-like structure~\cite{Crespi1996, Terrones1,
PCCP_planar_2, Lusk2008, Lusk3, Lusk5} consisting of non-hexagonal
carbon rings ~\cite{McEuen-graphene-quilts11}. Pure haeckelite
structures and their hybrids with graphene have a significantly
lower thermal conductivity than pure graphene~\cite{DT228}. Only
few theoretical studies have investigated electronic transport in
selected graphitic carbon nanostructures with non-hexagonal rings
including hybrid graphene-haeckelite
structures~\cite{Botello-Mendez09, NanoLett-Terrones04} and
haeckelite nanotubes~\cite{Lisenkov, transport_Li, Popovic}

Here we combine {\em ab initio} density functional theory (DFT)
structural studies with DFT-based nonequilibrium Green function
calculations to investigate how the presence of non-hexagonal
rings affects electronic transport in graphitic structures. We
find that infinite monolayers, finite-width nanoribbons and
nanotubes formed of 5-8 haeckelite with only 5- and 8-membered
rings are generally more conductive than their graphene-based
counterparts. Presence of haeckelite defect lines in the perfect
graphitic structure, a model of grain boundaries in CVD-grown
graphene, increases the electronic conductivity and renders it
highly anisotropic.

Haeckelites~\cite{Crespi1996,Terrones1,PCCP_planar_2,
Lusk2008,Lusk3,Lusk5} consist of periodic 2D arrangements of
non-hexagonal rings of $sp^2$ bonded carbon atoms. Even though
these structures have not been synthesized yet on a large scale,
similar atomic arrangements have been observed {\em (i)} in 5-7
and 5-8 defect lines forming the in-plane interface between
adjacent graphene
flakes~\cite{{McEuen-graphene-quilts11},{Lahiri10},{huang2011}},
{\em (ii)} in a vitreous atomic network formed during
electron-beam irradiation of graphene~\cite{glass}, and {\em
(iii)} in 5-7 ring structures filling graphene nanoholes during
the healing process of these defects~\cite{reknit}. Most
theoretical studies have focused on the equilibrium structure,
stability and growth stability of haeckelites~\cite{%
{Crespi1996},{Terrones1},{PCCP_planar_2},{Lusk2008},{Lusk3},{Lusk5},
{NanoLett-Terrones04},{Irle11},{graphene-allotropes},{Su2013},
{Sheng12},{yuliu}} and found these systems to be either metallic,
semi-metallic or semiconducting~\cite{Lusk3,Lusk5,Su2013}.

With continuing interest in CVD-grown graphene as an electronic
material, increased attention must be given to carrier scattering
at haeckelite-like grain boundaries connecting defect-free
graphene regions. The most plausible model geometry to investigate
charge transport in polycrystalline graphene is that of
interconnected strips of haeckelite and graphene. A consistent
picture should be obtained by comparing the effect of different
structural arrangements and widths of haeckelite and graphene
strips on the conductance and its anisotropy. As a counterpart to
transport studies in graphene nanoribbons and nanotubes, we
present corresponding results for haeckelite nanoribbons and
nanotubes.

\section{Methods}
%==================================================================

To gain insight into the equilibrium structure, stability and
electronic properties of haeckelite structures, we performed DFT
calculations as implemented in the \textsc{SIESTA}
code~\cite{SIESTA}. Infinite 2D layers and 1D ribbons and
nanotubes were separated by 10~{\AA} thick vacuum regions in a 3D
periodic arrangement. We used the
Ceperley-Alder~\cite{Ceperley1980} exchange-correlation functional
as parameterized by Perdew and Zunger~\cite{Perdew81},
norm-conserving Troullier-Martins
pseudopotentials~\cite{Troullier91}, and a double-$\zeta$ basis
including polarization orbitals. The reciprocal space was sampled
by a fine grid~\cite{Monkhorst-Pack76} of at least
$8{\times}8{\times}1$~$k$-points in the Brillouin zone of the 2D
primitive unit cell and its equivalent for 1D structures or larger
2D supercells. We used a mesh cutoff energy of $180$~Ry to
determine the self-consistent charge density, which provided us
with a precision in total energy of ${\lesssim}2$~meV/atom. All
geometries have been optimized using the conjugate gradient
method~\cite{CGmethod}, until none of the residual
Hellmann-Feynman forces exceeded $10^{-2}$~eV/{\AA}.

Electronic transport properties were investigated using the
nonequilibrium Green's function (NEGF) approach as implemented in
the \textsc{TRAN-SIESTA} code~\cite{transiesta}. Ballistic
transport calculations for optimized structures were performed
using a single-$\zeta$ basis with polarization orbitals, a
$180$~Ry mesh cutoff energy, and the same $k$-point
grid~\cite{Monkhorst-Pack76} as used for structure optimization.

%==================================================================
\section{Results and discussion}
%==================================================================

\subsection{5-8 haeckelite}

\begin{figure}
%\begin{figure}[!tb]
\includegraphics[width=1.0\columnwidth]{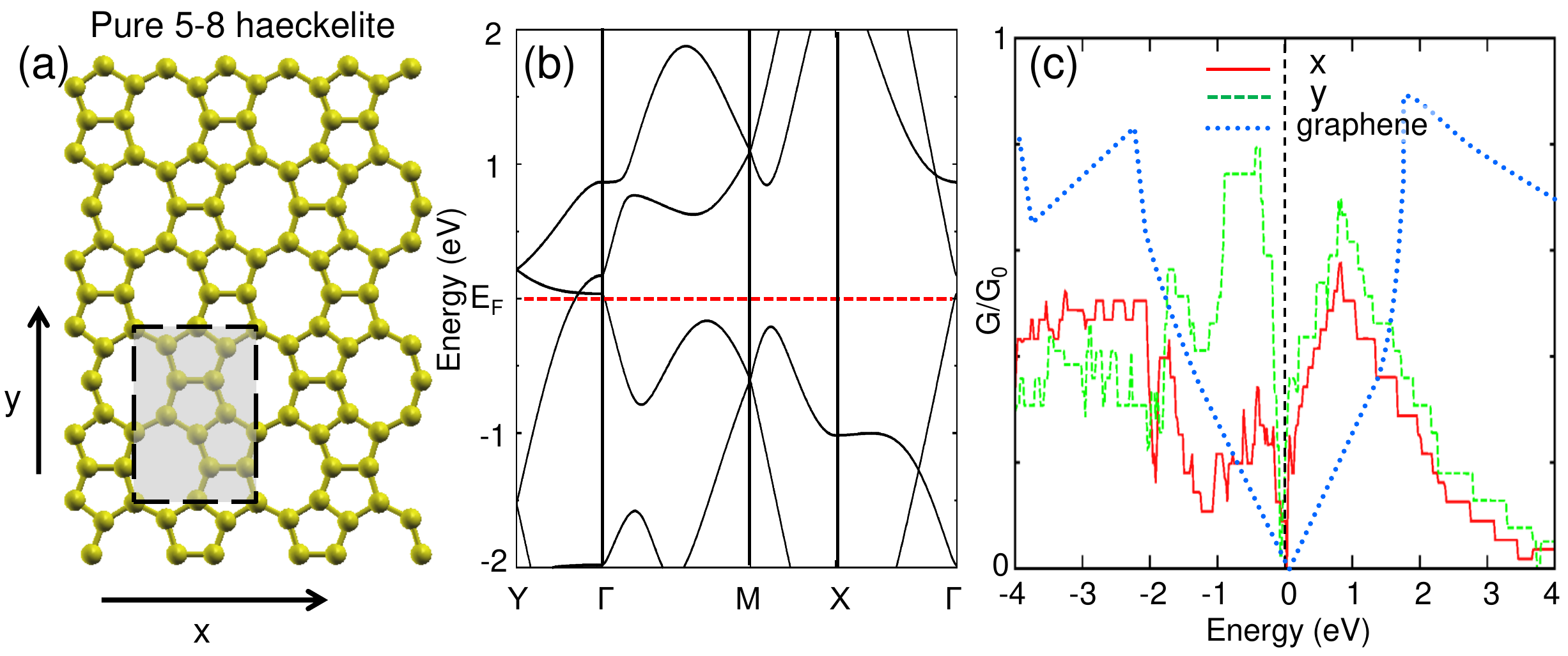}
\caption{(Color online) (a) Atomic structure and (b) electronic
band structure of 5-8 haeckelite with only 5- and 8-membered
rings. The primitive unit cell is indicated by the shaded region.
(c) Electronic conductance $G/G_{0}$ along the horizontal
$x$-direction, shown by the solid red line, and the vertical
$y$-direction, shown by the dashed green line. $G$ is normalized
by the width of the unit cell normal to the transport direction.
$E=0$ corresponds to carrier injection at $E_F$.} \label{fig:Fig1}
\end{figure}

% description of Figure 1
%     lattice: a1= 4.869 \AA and a2 = 6.927 \AA
A perfect 2D monolayer of 5-8 haeckelite containing only 5- and
8-membered rings is shown in Fig.~\ref{fig:Fig1}(a). The optimum
rectangular unit cell is spanned by the Bravais lattice vectors
$a_1$=4.87~{\AA} in the $x$-direction and $a_2$=6.93~{\AA} in the
$y$-direction. The 5-8 haeckelite structure is about 0.36~eV/atom
less stable than graphene, which is comparable to the stability of
narrow carbon nanotubes. Due to this relatively high stability, we
expect 5-8 haeckelite structures to coexists with graphene at
grain boundaries.

The electronic band structure of 5-8 haeckelite is presented in
Fig.~\ref{fig:Fig1}(b). In contrast to semimetallic graphene, 5-8
haeckelite is metallic and has a finite electronic density of
states at the Fermi level. Whereas the Fermi surface of graphene
consists of 6 isolated $k$-points, that of 5-8 haeckelite is a
line of finite length that intersects the $\Gamma$-Y high-symmetry
line, as seen in Fig.~\ref{fig:Fig1}(b).

%===========< FIGURE 1 >=================================================
%===========< FIGURE 1 >=================================================

%===========< FIGURE 2 >=================================================
\begin{figure}
%\begin{figure}[!tb]
\includegraphics[width=1.0\columnwidth]{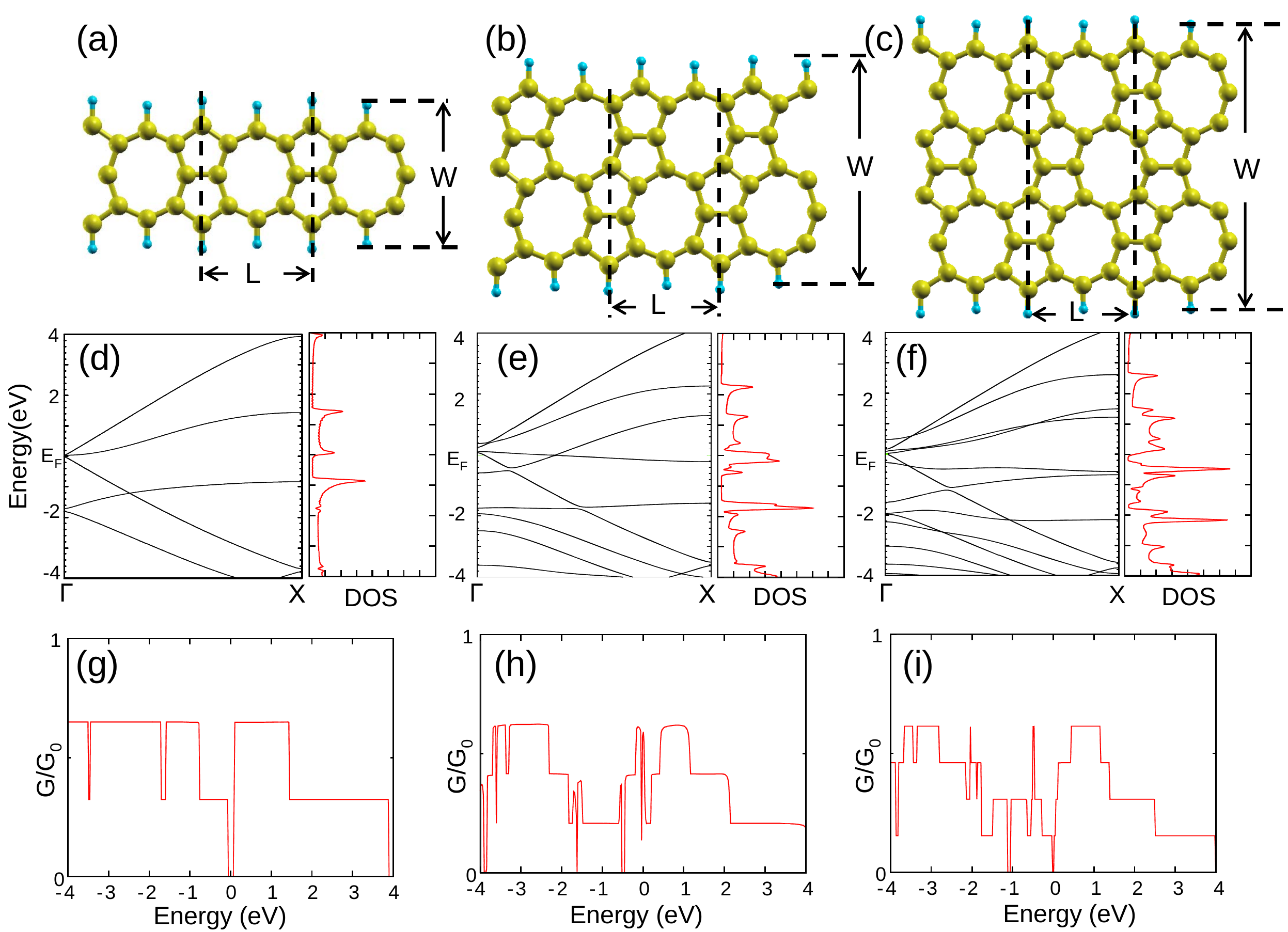}
\caption{(Color online) Atomic structure, electronic band
structure and transport properties of haeckelite nanoribbons
passivated by hydrogen at the edge. Atomic structure of haeckelite
nanoribbons with various width are shown in (a), (b), (c). The
corresponding electronic band structure and density of states is
shown in (d), (e), (f), and ballistic transport conductance
$G/G_0$ in (g), (h), (i). The density of state is given per atom.}
\label{fig:Fig2}
\end{figure}
%===========< FIGURE 2 >=================================================

%===========< FIGURE 3 >=================================================
\begin{figure}
%\begin{figure}[!tb]
\includegraphics[width=1.0\columnwidth]{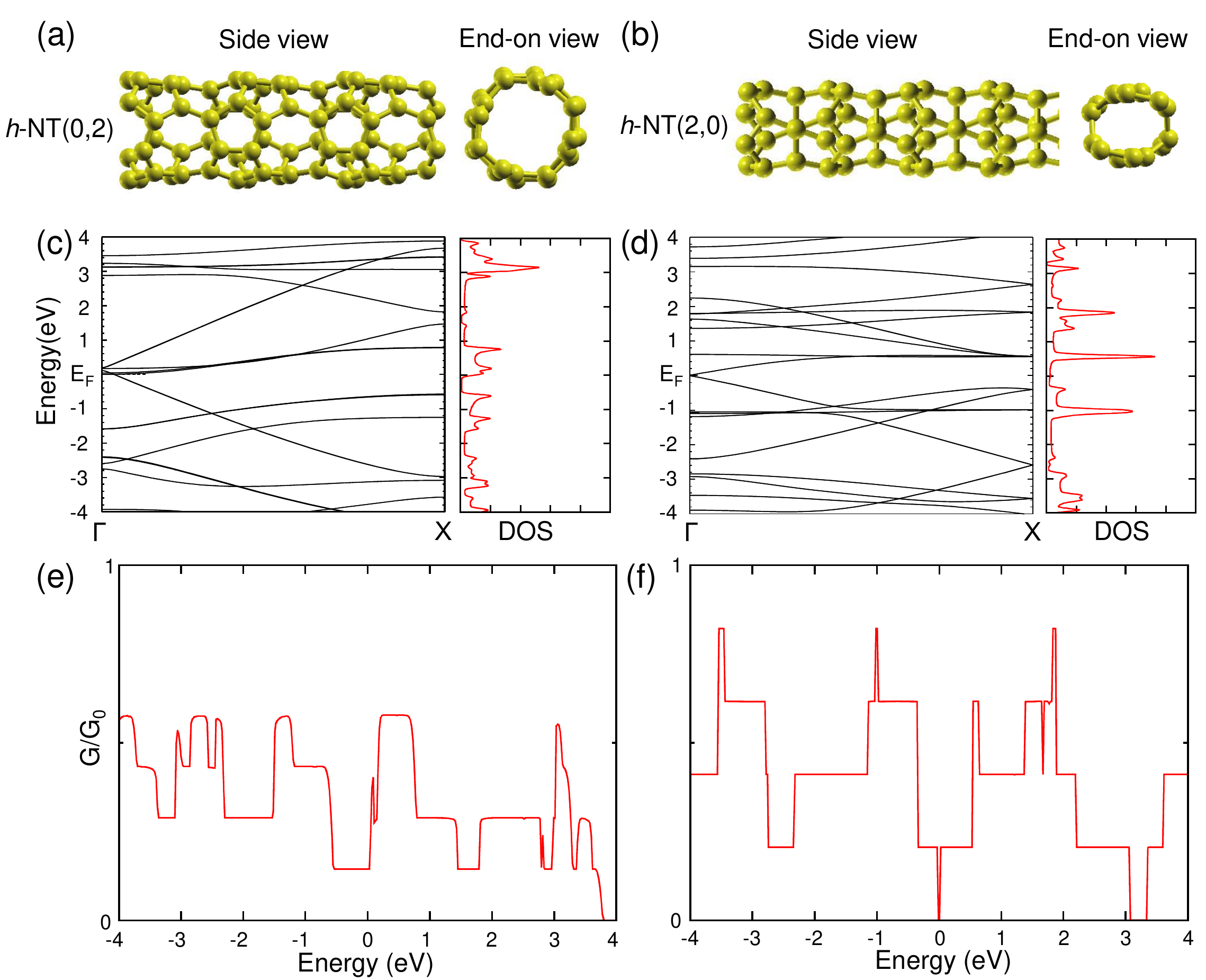}
\caption{(Color online) Structural and electronic properties of
haeckelite nanotubes. Atomic structure of (a) h-NT(0,2) and (b)
h-NT(2,0) nanotubes. The length of the unit cells is denoted by
$L$ and the width by $W$. The corresponding electronic band
structure and density of states is shown in (c), (d), and
ballistic transport conductance $G/G_0$ in (e), (f). The density
of state is given per atom.} \label{fig:Fig3}
\end{figure}
%===========< FIGURE 3 >=================================================

%===========< FIGURE 4 >=================================================
\begin{figure}
%\begin{figure}[!tb]
\includegraphics[width=1.0\columnwidth]{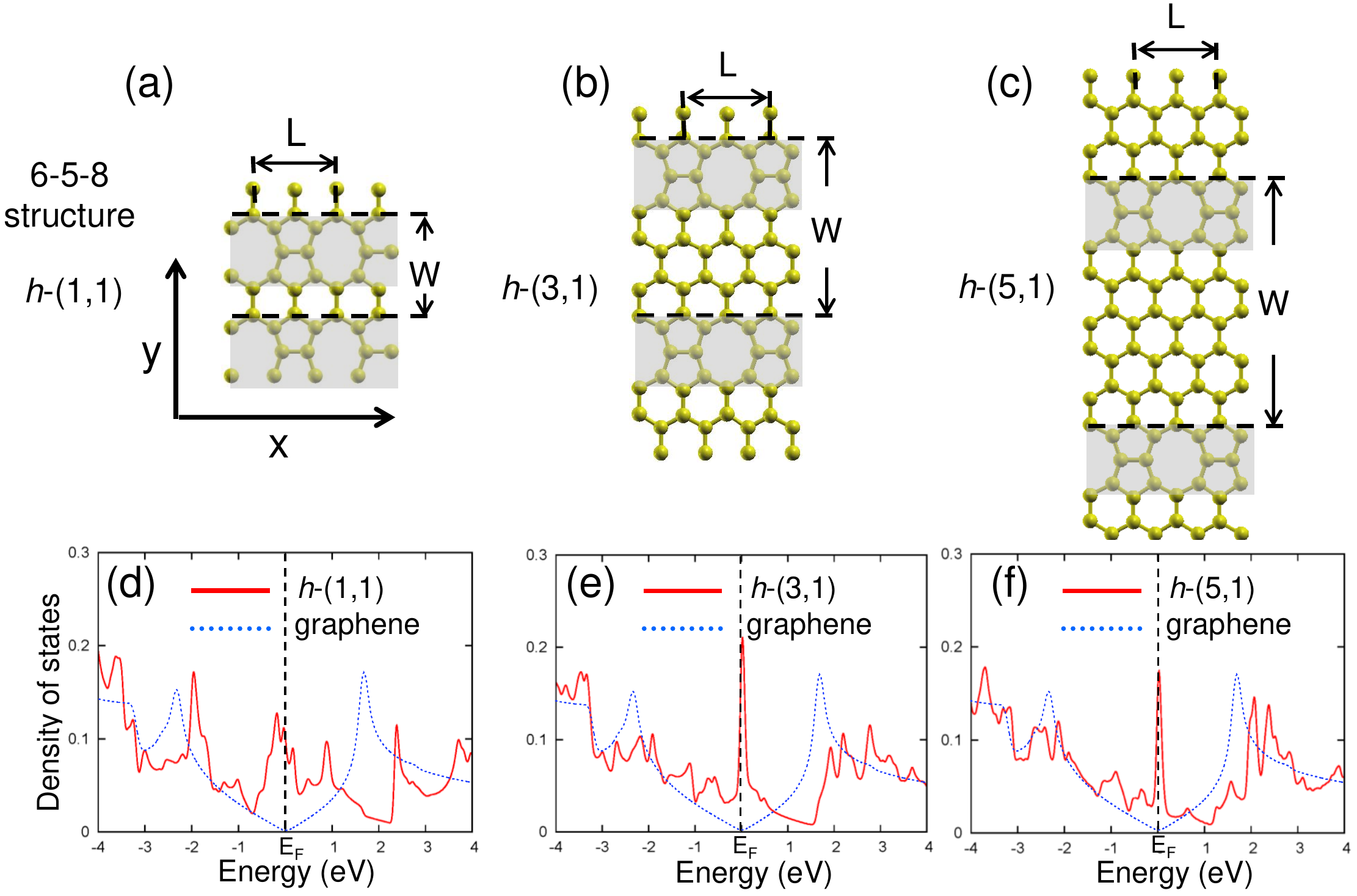}
\caption{(Color online) Structure and electronic properties of
hybrid haeckelite/graphene structures h-(n,m) consisting of
graphene strips interconnected by 5-8 haeckelite strips, with $n$
refereeing to the width of graphene and $m$ to that of haeckelite.
The upper panels depict the atomic structure of (a) h-(1,1), (b)
h-(3,1) and (c) h-(5,1) hybrid structures with $m=1$. The
corresponding electronic densities of states are shown in the
lower panels (d), (e) and (f). The length of the unit cells is
denoted by $L$ and the width by $W$. The density of state is given
per atom.} \label{fig:Fig4}
\end{figure}
%===========< FIGURE 4 >=================================================

%===========< FIGURE 5 >=================================================
\begin{figure}
\includegraphics[width=1.0\columnwidth]{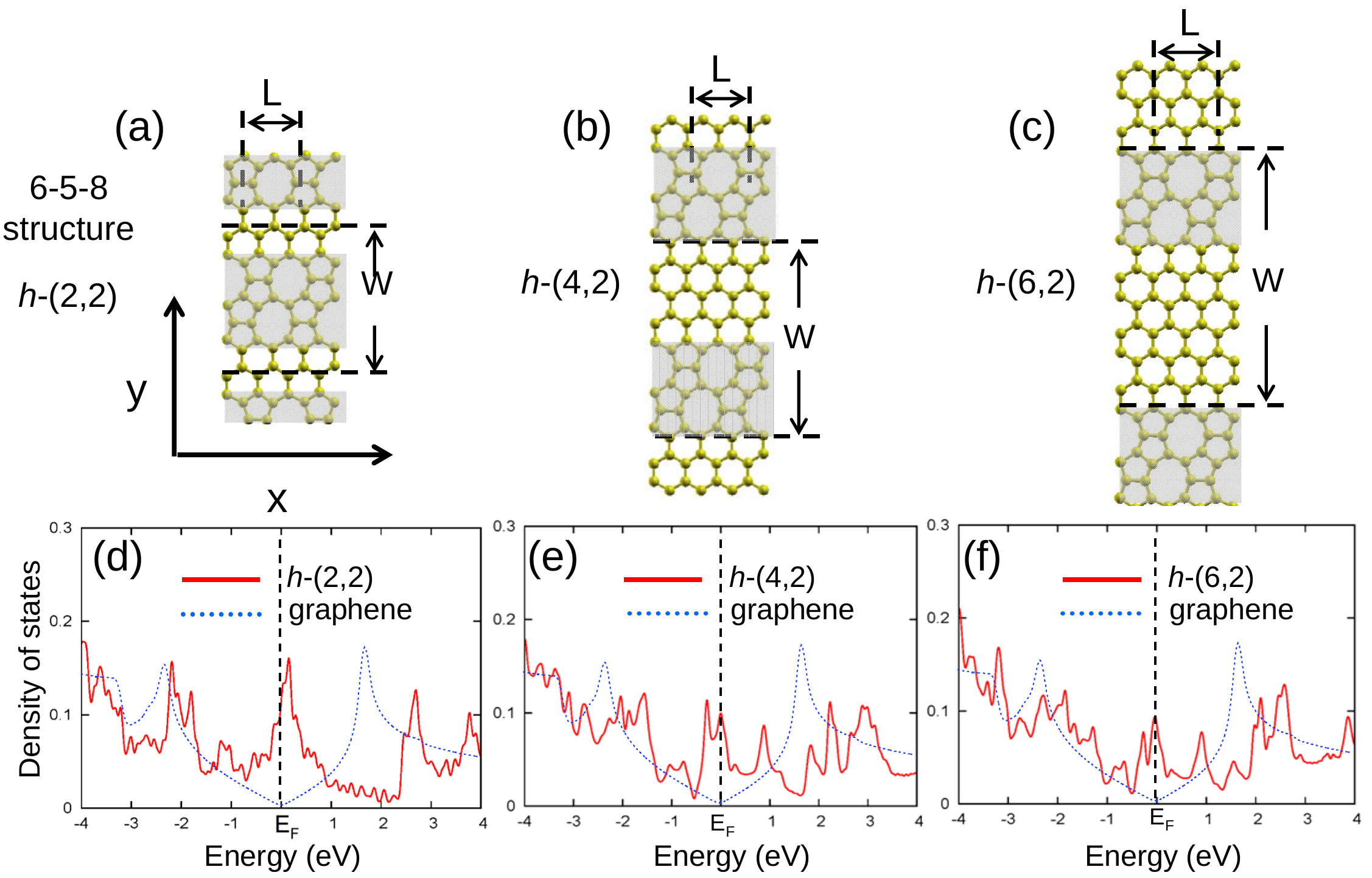}
\caption{(Color online) Structure and electronic properties of
hybrid haeckelite/graphene structures h-(n,m) consisting of
graphene strips interconnected by 5-8 haeckelite strips, with $n$
refereeing to the width of graphene and $m$ to that of haeckelite.
The upper panels depict the atomic structure of (a) h-(2,2), (b)
h-(4,2) and (c) h-(6,2) hybrid structures with $m=2$. The
corresponding electronic densities of states are shown in the
lower panels (d), (e) and (f). The length of the unit cells is
denoted by $L$ and the width by $W$. The density of state is given
per atom.} \label{fig:Fig5}
\end{figure}
%===========< FIGURE 5 >=================================================

%===========< FIGURE 6 >=================================================
\begin{figure}
\includegraphics[width=1.0\columnwidth]{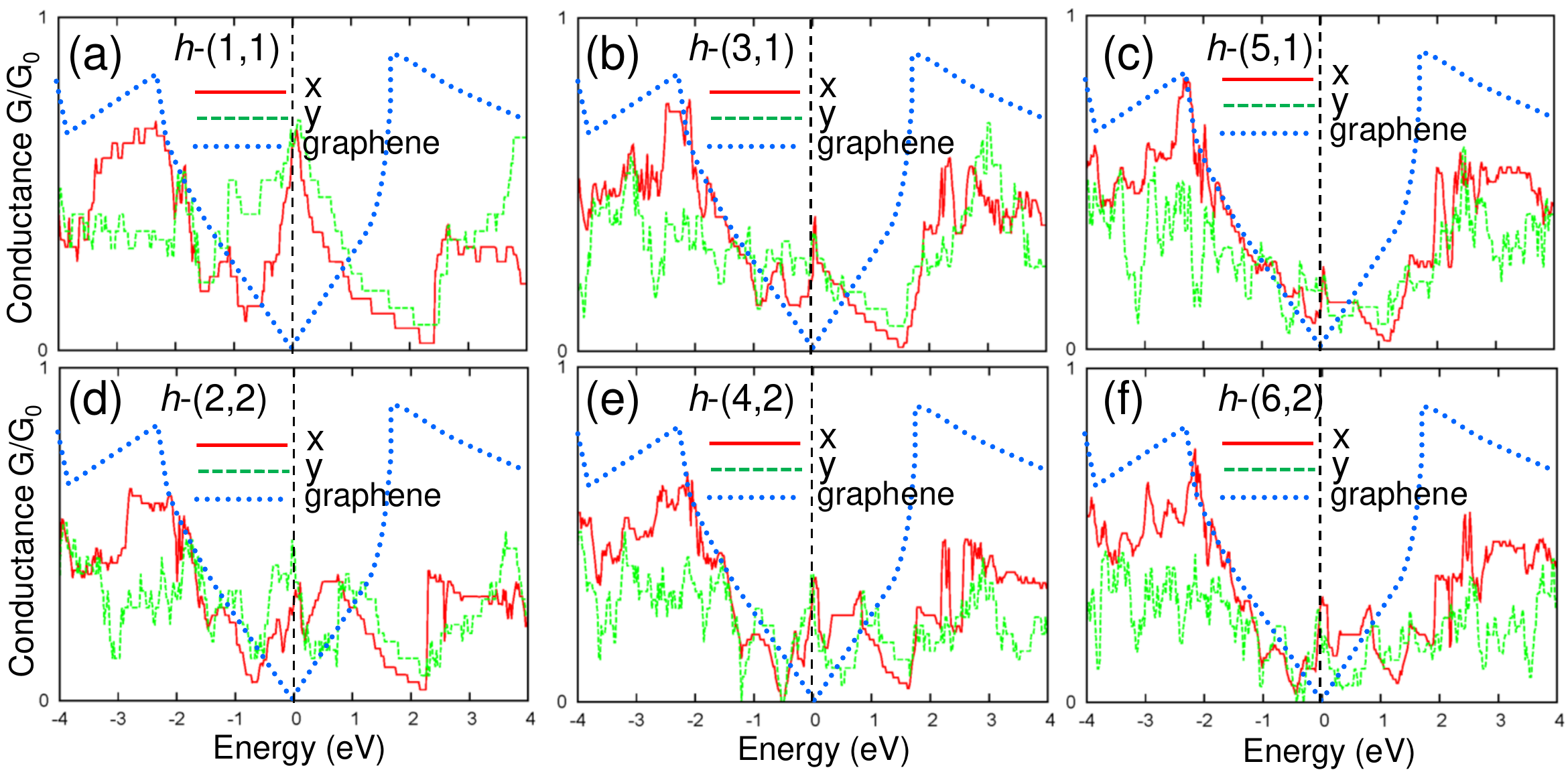}
\caption{(Color online) Electronic conductance $G$ of hybrid
haeckelite/graphene structures h-(n,m) consisting of graphene
strips interconnected by 5-8 haeckelite strips. Results are
presented for (a) h-(1,1), (b) h-(3,1), (c) h-(5,1), (d) h-(2,2),
(e) h-(4,2), (f) h-(6,2) with different widths $n$ of the graphene
strips and $m$ of the haeckelite strips. The structures and
transport directions are defined in
Figs.~\protect\ref{fig:Fig4}(a-c) and \protect\ref{fig:Fig5}(a-c).
$G$ is normalized by the width of the unit cell normal to the
transport direction. $E=0$ corresponds to carrier injection at
$E_F$. } \label{fig:Fig6}
\end{figure}
%===========< FIGURE 6 >=================================================

The results of our quantum transport calculation for this system
are shown in Fig.~\ref{fig:Fig1}(c). Besides the improved
conductivity over graphene, suggested by the increased density of
states at the Fermi level, we find the conductivity to be also
anisotropic, as expected when considering the atomic arrangement
in Fig.~\ref{fig:Fig1}(a). We find the electrical conductance
along the $y$ direction to be much higher than along the
$x$-direction, and even observe a very narrow transport gap at
$E_F$. These findings are consistent with a very anisotropic Fermi
surface that crosses the $\Gamma$-Y, but not the $\Gamma$-X
high-symmetry line in Fig.~\ref{fig:Fig1}(b).

\subsection{5-8 haeckelite nanoribbons and nanotubes}
%    (b) nano-ribbon with H termination
%          mono-strip nanoribbon: one 5-8 line
%               10 C atoms and 4 H atoms
%                lattice: z = 4.897 \AA (the periodic direction)
%               electronic property: semiconductor with 0.05 eV band gap
%          double-strip nanoribbon: two 5-8 lines
%               16 C atoms and 4 H atoms
%                lattice: z = 4.898 \AA (the periodic direction)
%               electronic property: metallic
%          triple-strip nanoribbon: two 5-8 lines
%               22 C atoms and 4 H atoms
%                lattice: z = 4.899 \AA (the periodic direction)
%               electronic property: semiconductor 0.02 eV band gap

Finite-width graphene nanoribbons and carbon nanotubes have
received wide attention, since -- unlike infinite graphene
monolayers -- some of these structures display sizeable band gaps.
In analogy to these structures, we also studied quantum transport
in 1D haeckelite nanoribbons (h-NRs) and nanotubes(h-NTs). In
Fig.~\ref{fig:Fig2}, we present our results for h-NRs with
different widths $W$ that are passivated by hydrogen at the edge.
The atomic structure of the three narrowest haeckelite nanoribbons
are shown in Figs.~\ref{fig:Fig2}(a-c). We find the optimum
lattice constant $L{\approx}4.9$~{\AA} to be nearly independent of
the width $W$. We also note that structures in
Fig.~\ref{fig:Fig2}(a) and \ref{fig:Fig2}(c) have mirror symmetry,
whereas that in Fig.~\ref{fig:Fig2}(b) does not.

As seen in panels (d-i) of Fig.~\ref{fig:Fig2}, we found the
electronic structure and conductance of h-NRs to depend
sensitively on the ribbon width. Structures with an odd number of
8-membered rings in the unit cell, depicted in
Figs.~\ref{fig:Fig2}(a) and \ref{fig:Fig2}(c), are narrow-gap
semiconductors, with the fundamental band gap decreasing with
increasing width from $E_g=0.05$~eV in Fig.~\ref{fig:Fig2}(a,d) to
$E_g=0.02$~eV in Fig.~\ref{fig:Fig2}(c,f). Structures with an even
number of 8-membered rings in the unit cell are all metallic. One
example with two 8-membered rings in the unit cell is shown in
Fig.~\ref{fig:Fig2}(b,e). Quantum conductance $G$ of the three
h-NRs in units of the conductance quantum $G_0$ is shown in
Fig.~\ref{fig:Fig2}(g-i). Of most interest is the presence or
absence of a transport gap at $E=0$, corresponding to carrier
injection at $E_F$. As expected, the semiconducting nanoribbons
depicted in Figs.~\ref{fig:Fig2}(a) and \ref{fig:Fig2}(c), have
also a finite transport gap, seen in Figs.~\ref{fig:Fig2}(g) and
\ref{fig:Fig2}(i). The metallic nanoribbon in
Fig.~\ref{fig:Fig2}(b) does not have a transport gap at $E=0$
according to Fig.~\ref{fig:Fig2}(h).

%    (c) 5-8 haeckelite nanotubes
%         (0,2) tube:  24 atoms/unit cell
%              lattice:  z= 4.868 \AA
%              stability: -154.634 eV/atom
%              electronic property: semi-metallic
%         (0,4) tube: 48 atoms/unit cell
%              lattice: z= 4.880 \AA
%              stability: -154.834 eV/atom
%              electronic property: semi-metallic
%         (2,0) tube: 24 atoms/unit cell
%              lattice: z = 6.923 \AA
%              stability: -154.270 eV/atom
%              electronic property: tiny band gap < 0.1 eV
%         (4,0) tube: 48 atoms/unit cell
%             lattice: z = 6.949 \AA
%             stability: -154.736 eV/atom
%             electronic property:  semimetallic

Similar to graphene-based carbon nanotubes, we can construct 5-8
haeckelite-based nanotubes, and characterize them by the chiral
index $(n,m)$ in analogy to carbon nanotubes. We present the
structure and electronic properties of two representative 5-8
haeckelite nanotubes, h-NT(0,2) and h-NT(2,0), in
Fig.~\ref{fig:Fig3}. Both the side and the end-on view of these
nanotubes in Fig.~\ref{fig:Fig3}(a,b) indicates that their optimum
cross-section is not as round and their surface not as smooth as
that of their graphitic counterparts, owing to the presence of 5-
and 8-membered rings. We found the ultra-narrow h-NT(0,2) and
h-NT(2,0) nanotubes to be stable, but highly strained. The
stability of the narrower h-NT(2,0) is lower by $0.81$~eV/atom and
that of the wider h-NT(0,2) is lower by $0.45$~eV/atom with
respect to the planar haeckelite structure depicted in
Fig.~\ref{fig:Fig1}(a).

The electronic band structure and density of states of the
h-NT(0,2) nanotube, shown in Fig.~\ref{fig:Fig3}(c), indicates
that h-NT(0,2) is metallic and has a non-zero density of states at
the Fermi energy. As expected, also the calculated quantum
conductance, shown in Fig.~\ref{fig:Fig3}(e), indicates
non-vanishing quantum conductance at $E=0$. Quite different are
the electronic structure and quantum conductance results for the
h-NT(2,0) nanotube, shown in Fig.~\ref{fig:Fig3}(d) and
\ref{fig:Fig3}(f), which display a very small fundamental and
transport gap of $<0.1$~eV. Unlike in graphene-based carbon
nanotubes, there is no general rule to predict whether a given
h-NT nanotube should be metallic or semiconducting.

\subsection{Hybrid haeckelite-graphene structures}

As a model of haeckelite-like grain boundaries connecting graphene
grains in polycrystalline graphene samples, we construct hybrid
haeckelite-graphene structures consisting of strips of 5-8 h-NRs
of various width inter-connecting bare zigzag graphene nanoribbons
with different widths. The hybrid systems, identified as
h-$(n,m)$, are characterized by the number $m$ of 8-membered rings
per unit cell and the number $n$ of hexagonal rings across the
width of the unit cell. We explored two families of hybrid
structures with $m=1$ and $m=2$, but different values of $n$.
These studies let us explore how the density of 5-8 line defects
should affect the electronic and transport properties of the
hybrid structures.

The optimum atomic arrangement and electronic structure of
h-$(n,1)$ hybrids, with $n=1,3,5$, is shown in
Fig.~\ref{fig:Fig4}. We found all the structures to be metallic,
as indicated by the peak in electronic densities of states at the
Fermi energy in Figs.~\ref{fig:Fig4}(d-f). The peak is associated
with the 5-8 defects and decreases in strength with decreasing
fraction of these defects in the structure, as the graphene strips
become wider.

The corresponding results for h-$(n,2)$ hybrids, with wider
haeckelite strips ($m=2$) and also wider graphene strips
characterized by $n=2,4,6$, are shown in Fig.~\ref{fig:Fig5}.
Similar to the $m=1$ family presented in Fig.~\ref{fig:Fig4}, also
the h-$(n,2)$ structures are all metallic, but the peak in the
density of states peak near $E_F$ does not appear as sharp as in
the h-$(n,1)$ structures. Similar to h-$(n,1)$ hybrids, the effect
of the defect line diminished with increasing $n$.

For the sake of illustration, we included the density of states of
pristine graphene in the same energy range and observe that
presence of 5-8 haeckelite line defects increases the density of
states near the Fermi level, benefitting conductivity.

We note that the narrow haeckelite strips within the h-$(n,1)$
hybrid structures correspond to the free-standing, but
hydrogen-passivated nanoribbon presented in
Fig.~\ref{fig:Fig2}(a). Similarly, also the wider haeckelite
strips within the h-$(n,2)$ hybrid structures have a counterpart
in the free-standing, but hydrogen-passivated nanoribbon presented
in Fig.~\ref{fig:Fig2}(b). Comparison between the density of
states of the free-standing haeckelite nanoribbons in
Fig.~\ref{fig:Fig2}(d,e) and the densities of states in
Figs.~\ref{fig:Fig4}(d-f) and Figs.~\ref{fig:Fig5}(d-f) indicates
that the electronic structure of free-standing and embedded
nanoribbons is very different.

The increasing similarity of h-$(n,m)$ structures with increasing
width $n$ of the graphene strips and defect-free graphene is also
reflected in the quantum conductance of these systems, shown in
Fig.~\ref{fig:Fig6}. Unlike the density of states, quantum
conductance is anisotropic. Especially for large values of $n$,
transport along the defect lines will increasingly resemble that
of graphene superposed with that of the conducting defect lines
acting as quantum conductors or metal wires. We conclude that
lines of non-hexagonal rings at grain boundaries in
polycrystalline graphene should enhance the conductance of this
system over pristine semimetallic graphene.

%==================================================================
\section{Conclusions}
%==================================================================

%We have studied the electronic and transport properties of
%haeckelite with 5- and 8- membered rings and haecklite/graphene
%hybrids. Based on our calculation, such haeckelite and
%haeckelite/graphene hybrid structures are stable and could exist
%in the experiments. We find haeckelite and haeckelite/graphene
%hybrids show a metallic character, rather than semimetallic
%graphene. Introducing haeckelite defective lines into graphene
%could enhance its electrical conductance, similar to grain
%boundary in CVD grown graphene. Such line defects could act as
%conducting wires within the graphene matrix. We also find that the
%1D nanotube and nanoribbon derived from haeckelite shows a
%diminishing band gap due to quantum confinement effect.

In conclusion, we have combined {\em ab initio} density functional
theory (DFT) structural studies with DFT-based nonequilibrium
Green function calculations to study how the presence of
non-hexagonal rings affects electronic transport in graphitic
structures found in polycrystalline graphene. We found that
infinite monolayers, finite-width nanoribbons and nanotubes formed
of 5-8 haeckelite with only 5- and 8-membered rings are generally
more conductive than their graphene-based counterparts. Presence
of haeckelite defect lines in the perfect graphitic structure, a
model of grain boundaries in CVD-grown graphene, increases the
electronic conductivity and renders it highly anisotropic.

\begin{ack}
The authors acknowledge financial support from the National
Science Foundation Cooperative Agreement No. EEC-0832785, titled
``NSEC: Center for High-rate Nanomanufacturing''. Computational
resources have been provided by the Michigan State University High
Performance Computing Center.
\end{ack}

\section*{References}
%==================================================================
%\addcontentsline{toc}{chapter}{REFERENCES}
%\bibliographystyle{iopart-num}% your bst file here
%\bibliography{haeckelec15} %your bib file here
%\end{document}
%==================================================================

\providecommand{\newblock}{}

\end{document}